\documentclass[preprint,12pt]{elsarticle}

\usepackage{graphicx} %% for loading postscript figures

\usepackage{amssymb}
\journal{Physica A}

\begin{document}

\begin{frontmatter}

\title{
Dynamics of stainless steel turning: Analysis by flicker-noise spectroscopy
}

\author[1]{Grzegorz Litak}
\author[2,5]{Yuriy S. Polyakov\corref{cor1}}
\ead{ypolyakov@uspolyresearch.com}
\author[3,4]{Serge F. Timashev}
\author[1]{Rafa\l{} Rusinek}

\cortext[cor1]{Corresponding author at: USPolyResearch, 906 Spruce St., Ashland, PA, 17921. Tel.: 347-673-7747.}

\address[1]{Department of Applied Mechanics, Lublin University of Technology, PL-20-618 Lublin, Poland}
\address[2]{USPolyResearch, 906 Spruce St., Ashland, PA 17921, USA}
\address[5]{Department of Computer Science, New Jersey Institute of Technology, Newark, NJ 07901, USA}
\address[3]{Karpov Institute of Physical Chemistry, ul. Vorontsovo pole 10, Moscow, 103064 Russia}
\address[4]{Institute of Laser and Information Technologies, Russian Academy of Sciences, ul. Pionerskaya 2, Troitsk, 142092 Russia}

%%%%%%%%%%%%%%%%%%%%%%%%%%%%%%%%%%%%%%%%%%%%%%%%%%%%%%%%%%%%%%%%%%%%%%
\begin{abstract}

We use flicker-noise spectroscopy (FNS), a phenomenological method for the analysis of time and spatial series operating on structure functions and power spectrum estimates, to identify and study harmful chatter
 vibrations in a regenerative turning process. The 3D cutting force components experimentally measured during stainless steel turning are analyzed, and the parameters of
their stochastic dynamics are estimated. Our analysis shows that the system initially exhibiting
 regular vibrations associated with spindle rotation becomes unstable to high-frequency noisy oscillations (chatter)
at larger cutting depths. We suggest that the chatter may be attributed to frictional stick-and-slip interactions between the contact surfaces of cutting tool and workpiece. We compare our findings with previously reported results obtained by statistical, recurrence, multifractal, and wavelet methods. We discuss the potential of FNS in monitoring the turning process in manufacturing practice.

\end{abstract}

\begin{keyword}


Flicker-noise spectroscopy \sep Cutting dynamics \sep Turning process 
\sep Chatter \sep Time series analysis
\end{keyword}

\end{frontmatter}

% \linenumbers

\section{INTRODUCTION}
\label{1}

Turning is a single-point cutting process that is used to manufacture rotational parts by removing excess material \cite{Ste97}. The turning process generally requires a lathe (machine tool), workpiece, fixture, and cutting tool, which form a system of interacting components. The machine tool and other components must be designed and operated to withstand specific forces, such as cutting, thrust, or feed, without causing significant deflections or chatter vibrations during the machining \cite{Ste97,Altintas2000,Mae00,Che08,Sch08}. These harmful chatter vibrations can reduce precision and product quality or even damage the cutting tool, if their amplitude is relatively high. The instability of turning to chatter vibrations puts a major constraint on the development of faster and higher-precision manufacturing processes required by modern 
industry \cite{Altintas2000,Mae00,Che08,Sch08}. Therefore, a detailed understanding of system instabilities causing chatter vibrations is currently needed to advance the manufacturing science and practice.

 The occurrence of unexpected chatter in cutting processes was first noticed by Taylor \cite{Taylor1907} in the beginning of the 20$^{\textrm{th}}$ century. The first study on 
chatter vibrations, performed in 1946 by Arnold, dealt with the analysis of the self-sustained mechanism of these vibrations \cite{Arnold1946}.
The following studies showed the importance of time-delayed regenerative effects \cite{Tobias1958,Stepan2001,Fofana2003,Wang2006} and
structural
dynamics \cite{Tlusty1963,Merit1965,Pratt2001}. It was later demonstrated that dry 
friction and stick-and-slip motion may be essential factors leading to chatter vibrations \cite{Wu1985a,Wu1985b,Grabec1988}. Recently, chaotic and intermittent oscillations caused
by various system nonlinearities were theoretically predicted and experimentally detected 
\cite{Tansel1992,Gradisek1996,Wiercigroch1997a,Wiercigroch1997b,Gradisek1998a,Gradisek1998b,Marghitu2001,Litak2002,Warminski2003,Gradisek2002,Sen2007}. Stochastic mechanisms of chatter generation usually attributed to the initial workpiece surface roughness were also studied \cite{Wiercigroch1997b,Litak2004}.

Over the past decade, the fast development of cutting technology gave way to a
reliable high-speed cutting procedure for metals. But the progress is hindered by the requirement to eliminate and stabilize chatter vibrations arising in metal cutting  \cite{Altintas2000,Mae00,Che08}. Currently, the development of adaptive control procedures, based
on relatively short time series and effective for various metals and other materials, is one of the main goals of manufacturers. To this end, various methods of time series analysis are applied to study the onset of chatter vibrations and the dynamic characteristics of the cutting process. The dynamics of force components in the turning of steel CS45 (ISO) was examined for a range of cutting depths using several techniques of nonlinear time series analysis: surrogate data tests, correlation dimension and entropy rate estimates \cite{Gradisek1998a}. The authors identified two modes: chatter-free at lower values of cutting depth, corresponding to a linear random process, and chatter at higher values of cutting depth, possibly related to a low-dimensional weakly chaotic process. The dynamics of force components for stainless steel turning was studied by statistical (standard deviation, skewness, kurtosis, and root mean square) and recurrence (plots, recurrence quantification approximation) methods \cite{Litak2012}. It was shown that chatter vibrations become more pronounced as cutting depth increases. This result is also supported by another study where the same time series were analyzed using multifractal and wavelet methods \cite{Lit11}. 

        In the present paper, we study the dynamics of stainless steel turning by flicker-noise spectroscopy (FNS), a phenomenological approach to the analysis of time and spatial series operating on transient structure functions and power spectrum estimates \cite{Tim07a,Tim10a,Tim07b,Tim06a,Tim08a,Mir11a,Tim12}. We apply this method to the experimental time series for cutting forces to study different types of system dynamics. Namely, we vary a single system
parameter, cutting depth $h_d$, and observe qualitative changes in the system
dynamics, which makes it possible to optimize the manufacturing of rotational parts. We compare our findings with previously reported studies performed using statistical and nonlinear time series analysis methods.

The paper is structured as follows. In Section 2, we provide the fundamentals of FNS.
In Section 3, we briefly describe the experimental setup and then apply the FNS parameterization algorithm and cross-correlation function to the analysis of cutting force components in the time domain. Section 4 presents the concluding remarks.

\section{FLICKER-NOISE SPECTROSCOPY}
\label{2}

Here, we will only deal with the basic FNS relations needed to understand the parameterization procedure and cross-correlation function used in the analysis of the 
stainless steel turning time series. FNS is described in more detail elsewhere \cite{Tim07a,Tim10a,Tim07b,Tim06a,Tim08a}.

Flicker-noise spectroscopy is a phenomenological method for analyzing complex signals with stochastically varying components using Kolmogorov transient structure functions and autocorrelation-based estimates of power spectrum. It should be noted that the term ``stochastic", here and further in this paper, refers to random variability in the signals of complex systems characterized by nonlinear interactions, dissipation, and inertia \cite{Tim10a,Tim07a}. Conceptually, the method separates the analyzed signal $V(t)$, where $t$ is time, into three components: low-frequency regular component corresponding to system-specific ``resonances"  and their interferential (nonlinear) contributions, stochastic ``jump" (random walk) component at larger frequencies corresponding to a dissipative process of anomalous diffusion, and stochastic highest-frequency ``spike" component corresponding to inertial (non-dissipative) effects.

The parameterization is based on the analysis of the information contained in autocorrelation function
\begin{equation}
\psi (\tau) = \left\langle {V(t)V(t + \tau )} \right\rangle_{T-\tau}, \label{eq1}
\end{equation}
where $\tau$ is the time lag parameter ($0 < \tau \le T_M$), $T_M$ is the upper bound for $\tau$ ($T_M \le T/2$), and $T$ is the averaging window. The angular brackets in relation (\ref{eq1}) stand for the averaging over time interval $T-\tau$:  
\begin{equation}
\left\langle {(...)} \right\rangle_{T-\tau}  = {1 \over {T-\tau}}\int^{T-\tau}_{0} {(...) \,dt}. \label{eq2}
\end{equation}
The averaging over interval $T-\tau$ implies that all the characteristics that can be extracted by analyzing functions $\psi(\tau)$ should be regarded as the average values on this interval.
 
 To extract the information contained in $\psi (\tau )$ ( $\left\langle {V(t)} \right\rangle = 0$  is assumed), the following transforms, or "projections", of this 
function are analyzed:
\newline cosine transforms (``power spectrum" estimates) $S(f)$, where $f$ is the frequency,
\begin{equation}
S(f) = 2 \int^{T_M}_{0} { \left\langle {V(t)V(t + t_1 )} \right\rangle_{T-\tau} \, \cos({2 \pi f t_1}) \,dt_1} \label{eq3}
\end{equation}
\newline and its difference moments (Kolmogorov transient structure functions) of the second order $\Phi^{(2)} (\tau)$
%eq4
\begin{equation}
\Phi^{(2)} (\tau) = \left\langle {\left[ {V(t) - V(t+\tau )} \right]^2 } \right\rangle_{T-\tau}. \label{eq4}
\end{equation}

Here, we use the quotes for power spectrum because according to the Wiener-Khinchin theorem the cosine (Fourier) transform of autocorrelation function is equal to the power spectral density only for wide-sense stationary signals at infinite integration limits.

The information contents of $S(f)$ and $\Phi^{(2)}(\tau)$ are generally different, and the parameters for both functions are needed to solve parameterization problems. By considering the intermittent character of signals under study, interpolation expressions for the stochastic components $S_s(f)$ and ${\Phi_s}^{(2)} (\tau)$ of $S(f)$ and $\Phi^{(2)} (\tau)$, respectively, were derived using the theory of generalized functions by \citet{Tim06a}. It was shown that the stochastic components of structure functions $\Phi^{(2)} (\tau)$ are formed only by jump-like (random walk) irregularities, and stochastic components of functions $S(f)$, which characterize the ``energy side" of the process, are formed by spike-like (inertial) and jump-like irregularities. It should be noted that $\tau$ in Eqs. (\ref{eq1})-(\ref{eq4}) is considered as a macroscopic parameter exceeding the sampling period by at least one order of magnitude. This constraint is required to derive the expressions and separate out contributions of dissipative jump-like and inertial (non-dissipative) spike-like components. 

 The basic idea in parameterization is to use two three-parameter interpolation expressions for the stochastic components. The first 
 expression is used to determine the spectral contribution of the stochastic components of signal $V(t)$ and exclude the contribution of the 
 low-frequency (resonant) component to the parameters related to jump- and spike-like stochastic irregularities. The second interpolation, which deals 
 with the stochastic component in the structure function, is used to determine the parameters characterizing the series of jump-like 
 irregularities, which correspond to a dissipative process of anomalous diffusion \cite{Tim10a}.

 Six stochastic FNS parameters are introduced: $T_0$, the correlation time for jump- and spike-like irregularities after which the 
 self-similarity observed in power spectrum estimate breaks down; $S_s(T_0^{-1})$, the power spectrum estimate at frequency $T_0^{-1}$, which 
 accounts for the "intensity" of jump- and spike-like irregularities in the highest-frequency interval; $n$, the flicker-noise parameter, 
 which characterizes the rate of loss of correlations in the series of high-frequency irregularities in time intervals $T_0$; $\sigma$, the 
 standard deviation of the value of the measured dynamic variable from the slowly varying regular component, which is based solely on 
 jump-like irregularities; $T_1$, the correlation time for jump-like irregularities ("random walks") in stochastically varying signal $V(t)$; $H_1$, the 
  Hurst exponent (this estimate of the Hurst component is also referred to as the Hausdorff exponent \cite{Mal99}), which describes the rate at which the dynamic variable 
"forgets" its values on the time intervals that are less than $T_1$.

FNS interpolations were derived for a wide-sense stationary signal in which the phenomenological parameters are the same at each level of 
 the system evolution hierarchy. At the same time, they can also be applied to the analysis of real signals, which are generally 
  nonstationary, but can be characterized by a finite standard deviation within a specific averaging window. In this case, the real signals at specific averaging intervals 
and sampling frequencies should be regarded as 
 quasi-stationary with certain values of standard deviation and other phenomenological parameters. It should be noted that the values of phenomenological parameters may vary on 
 different quasi-stationary intervals. Sometimes, the signals under study are characterized by two or more scales. In this case, one can use interpolation expressions with 
higher numbers of parameters \cite{Tim07b,Pol12}.

The detailed FNS parameterization algorithm in discrete form is presented in Ref. \cite{Tim12}. 

The information about the dynamics of 
 correlations between variables $V_i(t)$ and $V_j(t)$, where indices $i$ and $j$ denote two different signals, can be extracted by analyzing the temporal 
 variations of various correlators. Here, we will limit our attention to the simplest two-signal correlation expression characterizing the links 
between $V_i(t)$ and $V_j(t)$ \cite{Tim07a,Tim07b,Tim06a}:

\begin{equation}
 q_{ij} (\tau, \theta _{ij} ) = \left\langle {\left[ {{{V_i (t) - V_i (t + \tau )} \over {\sqrt {\Phi _i^{\left( 2 \right)} \left( \tau  
 \right)} \,\,}}} \right]  \left[ {{{V_j (t + \theta _{ij} ) - V_j (t + \theta _{ij}  + \tau )} \over {\sqrt {\Phi _j^{\left( 2 \right)} 
 \left( \tau  \right)} }}} \right]} \right\rangle _{T - \tau  - \left| {\theta _{ij} } \right|},  \label{eq20}
\end{equation}
where $\tau$  is the time lag, $\theta_{ij}$ is the time shift parameter. Higher values of $\tau$ correspond to coarser (low-resolution) analysis of cross-correlations.

In discrete form, Eq.~(\ref{eq20}) is written in Ref. \cite{Tim12}. 

The cross-correlation expression $q_{ij} (\tau, \theta _{ij} )$ is a function of temporal parameters $\tau$ and $\theta_{ij}$, which can be represented as a three-dimensional plot. The dependence of cross-correlation $q_{ij}(\tau, \theta_{ij})$ on $\theta_{ij}$ describes the cause-and-effect relation
 between signals $V_i(t)$ and $V_j(t)$. When $\theta_{ij} > 0$, we look at the the changes in signal $V_i(t)$ that bring about changes in signal $V_j(t)$, when $\theta_{ij} < 0$, the opposite is true. The dependence of the value and magnitude of cross-correlation  $q_{ij}(\tau, \theta_{ij})$ on $\tau$ and $\theta_{ij}$ can be used to analyze the cause-and-effect dynamics with signals $V_i(t)$ and $V_j(t)$ 
changing in phase ($q_{ij} > 0$) and in antiphase ($q_{ij} < 0$). 

Of most interest for the analysis are the intervals of $\tau$ and $\theta_{ij}$ where the cross-correlation function $q_{ij}$ is closest to positive unity (maximum level of positive correlations) or negative unity (maximum level of negative correlations). 

The magnitude and behavior of the two-parameter expression (\ref{eq20}) may significantly depend on the value of selected averaging interval $T$ and upper-bound values of $\tau$ and $\theta_{ij}$, which we will refer to as $\tau_{\max}$ and $\theta_{\max}$. From the statistical reliability point of view, we set a constraint of $\tau_{\max} + |\theta_{\max}| \le T/2$.

\section{RESULTS AND DISCUSSION}
\label{3}

\subsection{Experimental data }
\label{3.1}

%fig1
\begin{figure}
\begin{center}
\includegraphics[width=8.0cm]{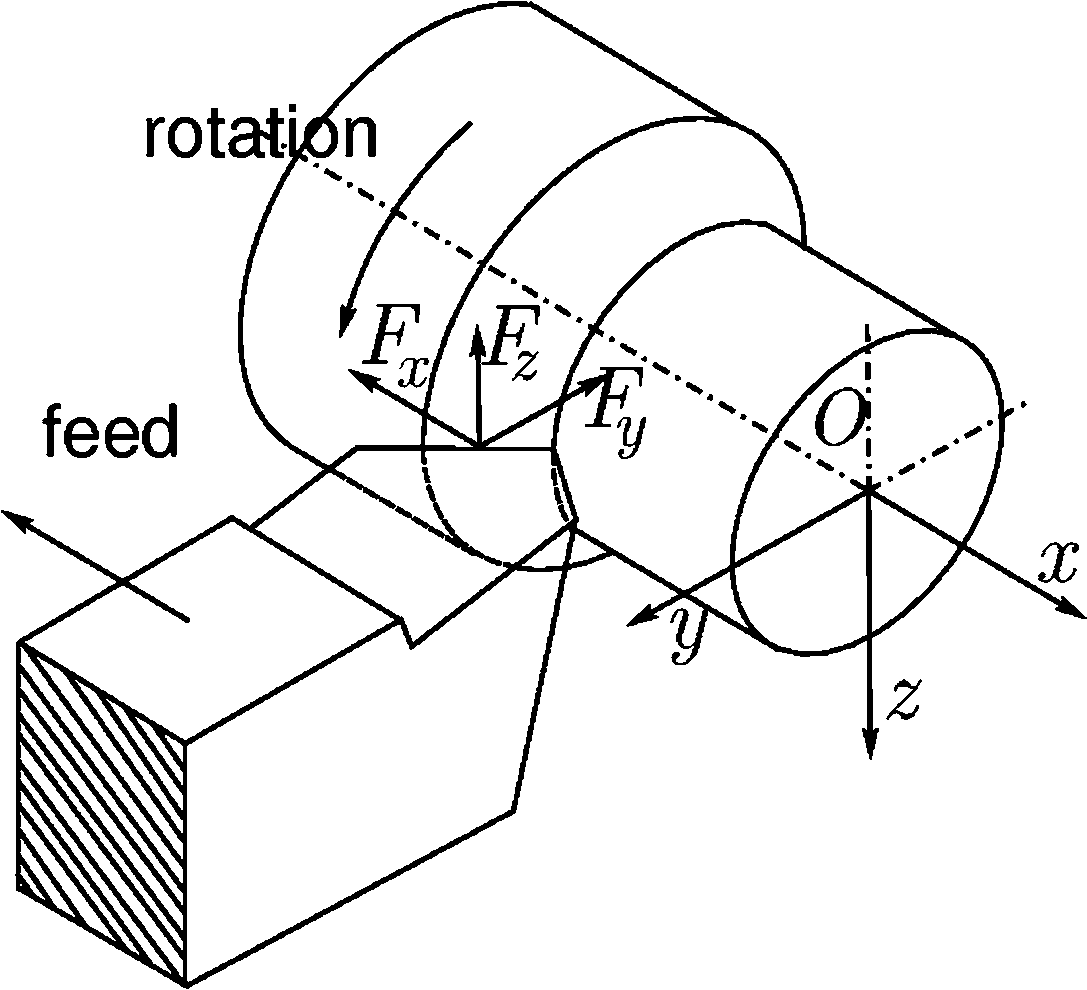}
\caption{Physical model of a regenerative turning process.  The
three measured orthogonal force components are feed $F_x$, thrust $F_y$, and cutting $F_z$.}
\end{center}
\end{figure}

Our turning experiments were conducted using a circular workpiece of stainless steel (EZ6NCT25) with a diameter of 22mm and tool
cutting edge angle of 45 degrees. The spindle angular velocity was fixed at 780
rpm while the corresponding feeding rate was 0.25 mm/rev. Three orthogonal force components were measured in each experiment by a piezoelectric dynamometer: feed $F_x$, thrust $F_y$, and cutting $F_z$ (Fig. 1). The measurements were registered with a sampling rate $f_d$ of 2 kHz and accuracy of 3\% for three different cutting depths $h_d$. Each measurement was recorded using a data acquisition computer card with 12-bit resolution. It should be noted that the actual cutting depth was not constant. Hence, the cutting depth $h_d$ of the workpiece should be considered as an average value.

%\section{RESULTS}

\subsection{Parameterization of cutting force time series }
\label{3.2}

%fig2
\begin{figure}
\begin{center}
\includegraphics[width=11.0cm]{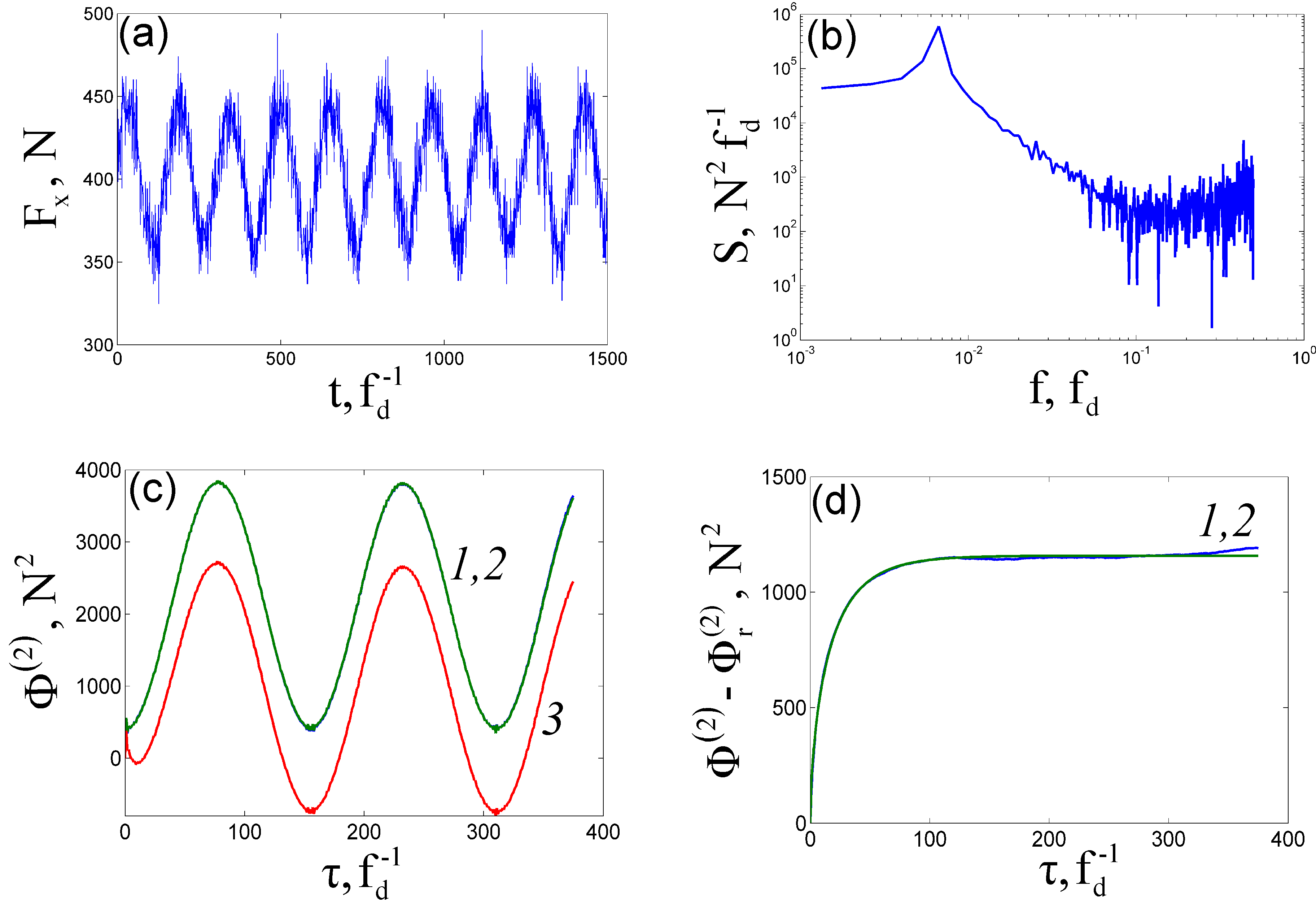}% Here is how to import EPS art
 \caption{\label{fig:1} (Color online) FNS parameterization of $F_x$ at $h_d = 1.00$ mm: (a) original time series;  (b) power spectrum estimate $S(f)$; 
 (c) structure functions $\Phi^{(2)}(\tau)$: 1, experimental; 2, calculated; 3, 
 contribution of resonant frequencies; (d) stochastic component of $\Phi^{(2)}(\tau)$: 1, experimental; 
2, calculated.}
\end{center}
\end{figure}

\begin{figure}
\begin{center}
\includegraphics[width=11.0cm]{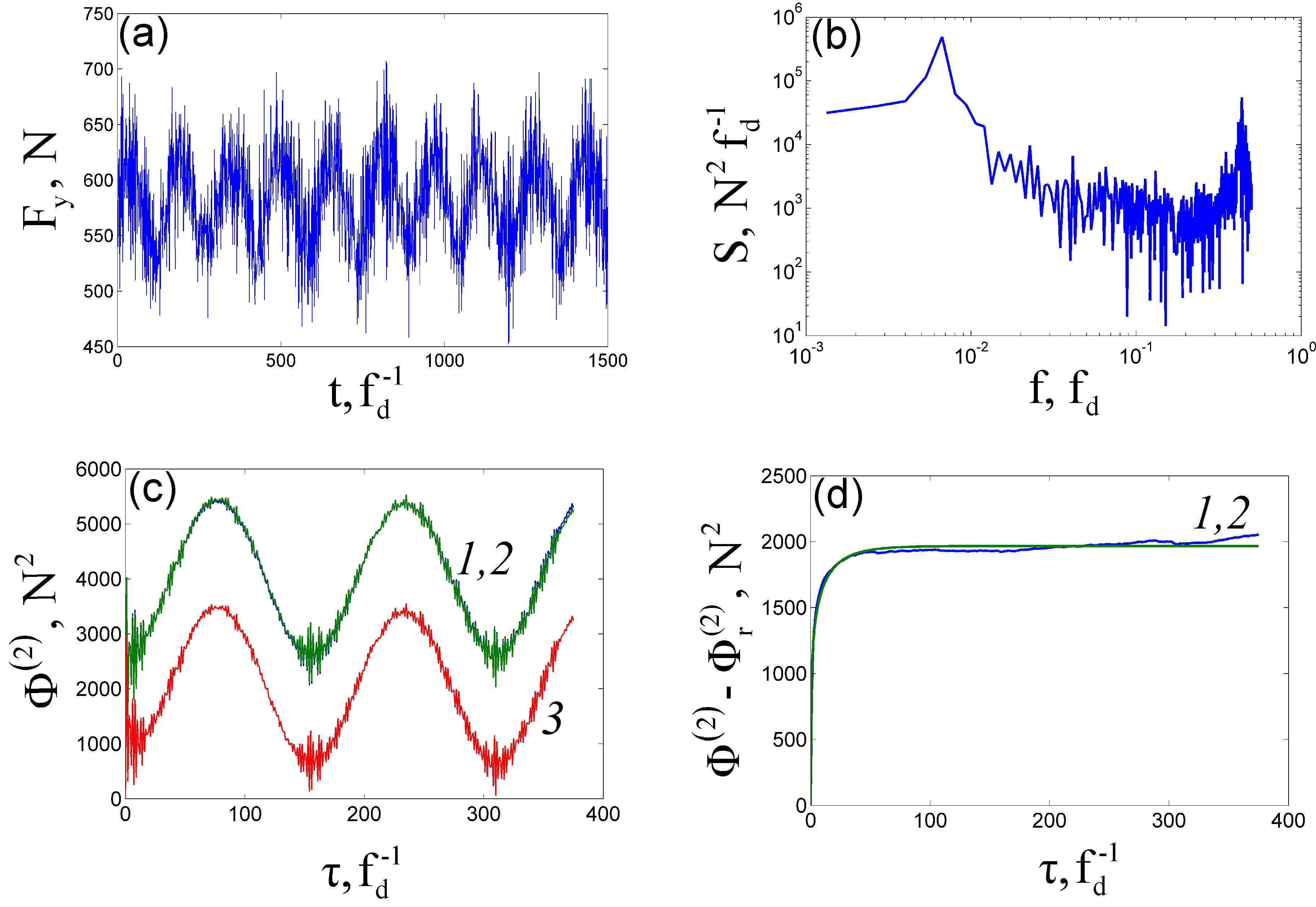}% Here is how to import EPS art
\caption{\label{fig:2} (Color online) FNS parameterization of $F_y$ at $h_d = 1.00$ mm: nomenclature as in Fig. 2.}
\end{center}
\end{figure}

\begin{figure}
\begin{center}
\includegraphics[width=11.0cm]{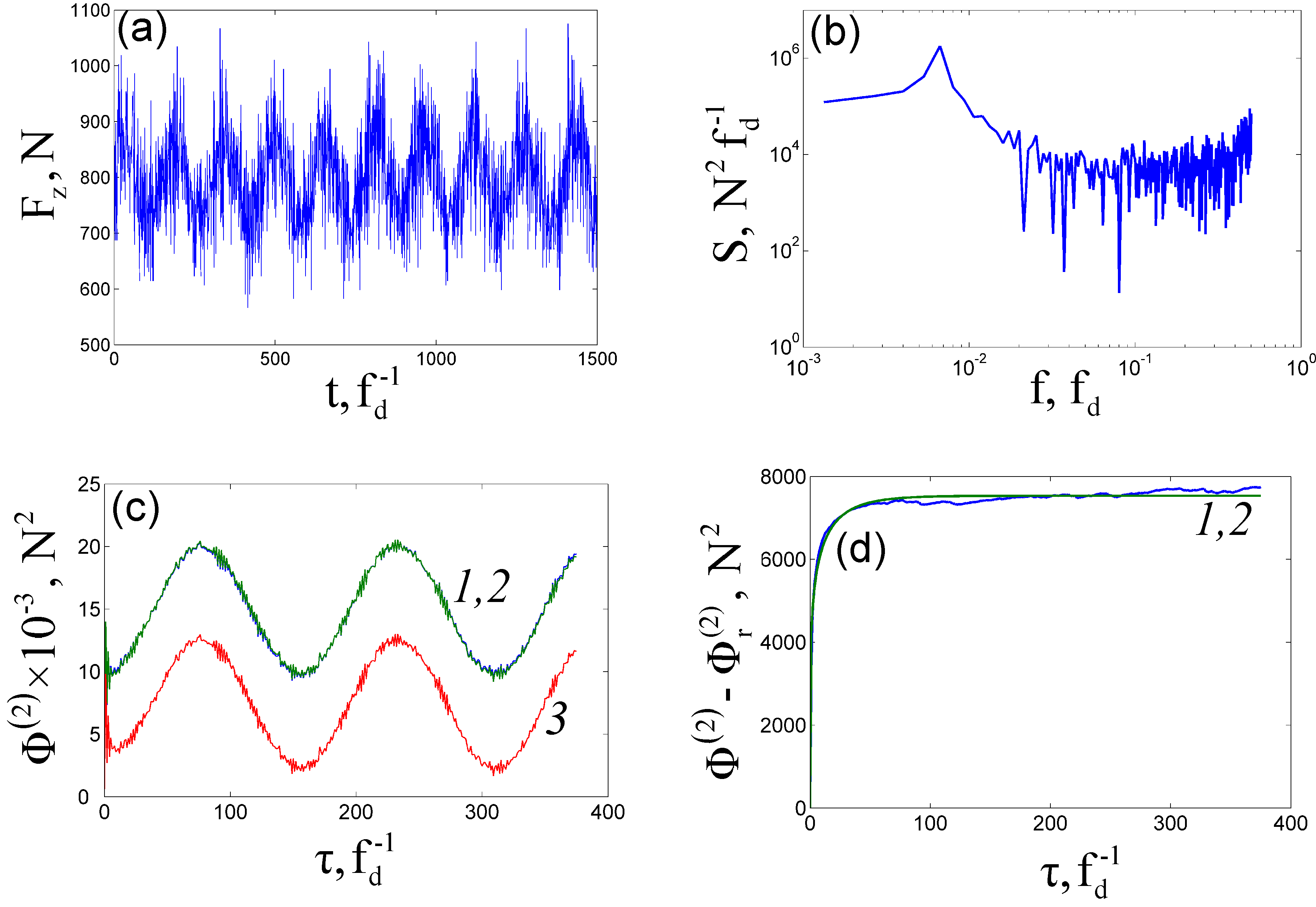}% Here is how to import EPS art
\caption{\label{fig:3} (Color online) FNS parameterization of $F_z$ at $h_d = 1.00$ mm: nomenclature as in Fig. 2.}
\end{center}
\end{figure}

\begin{figure}
\begin{center}
\includegraphics[width=11.0cm]{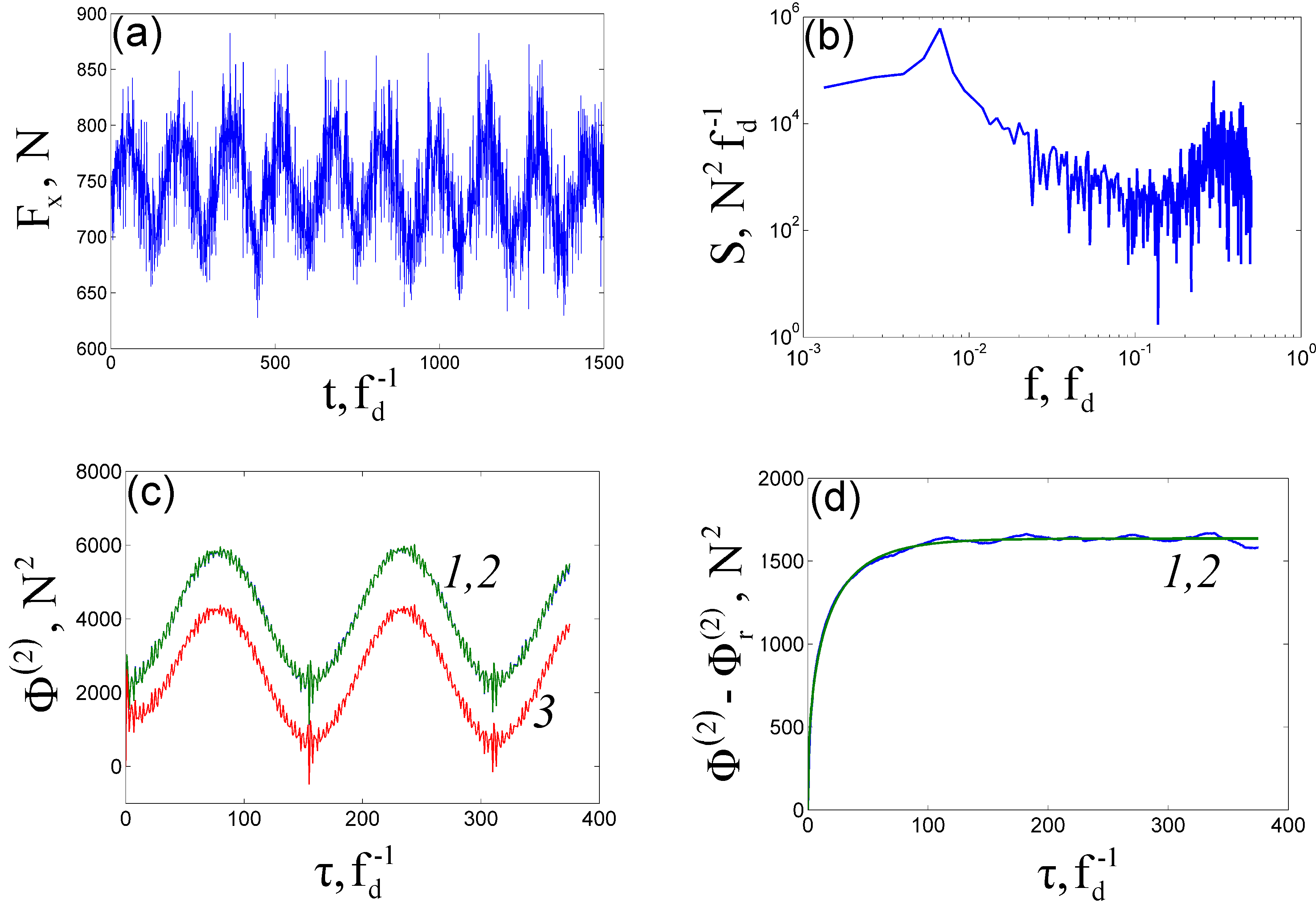}% Here is how to import EPS art
\caption{\label{fig:4} (Color online) FNS parameterization of $F_x$ at $h_d = 2.30$ mm: nomenclature as in Fig. 2.}
\end{center}
\end{figure}

\begin{figure}
\begin{center}
\includegraphics[width=11.0cm]{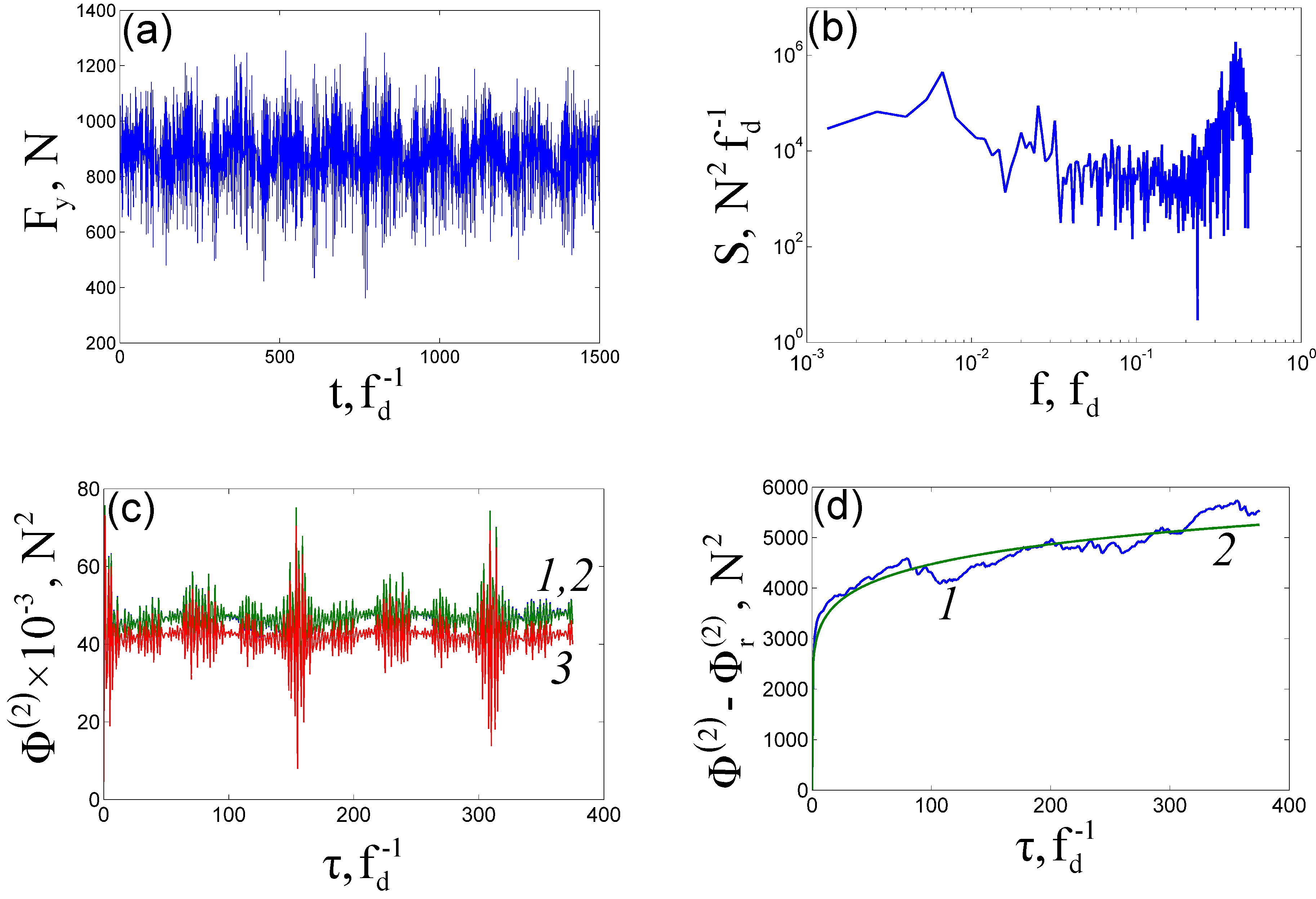}% Here is how to import EPS art
\caption{\label{fig:5} (Color online) FNS parameterization of $F_y$ at $h_d = 2.30$ mm: nomenclature as in Fig. 2.}
\end{center}
\end{figure}

\begin{figure}
\begin{center}
\includegraphics[width=11.0cm]{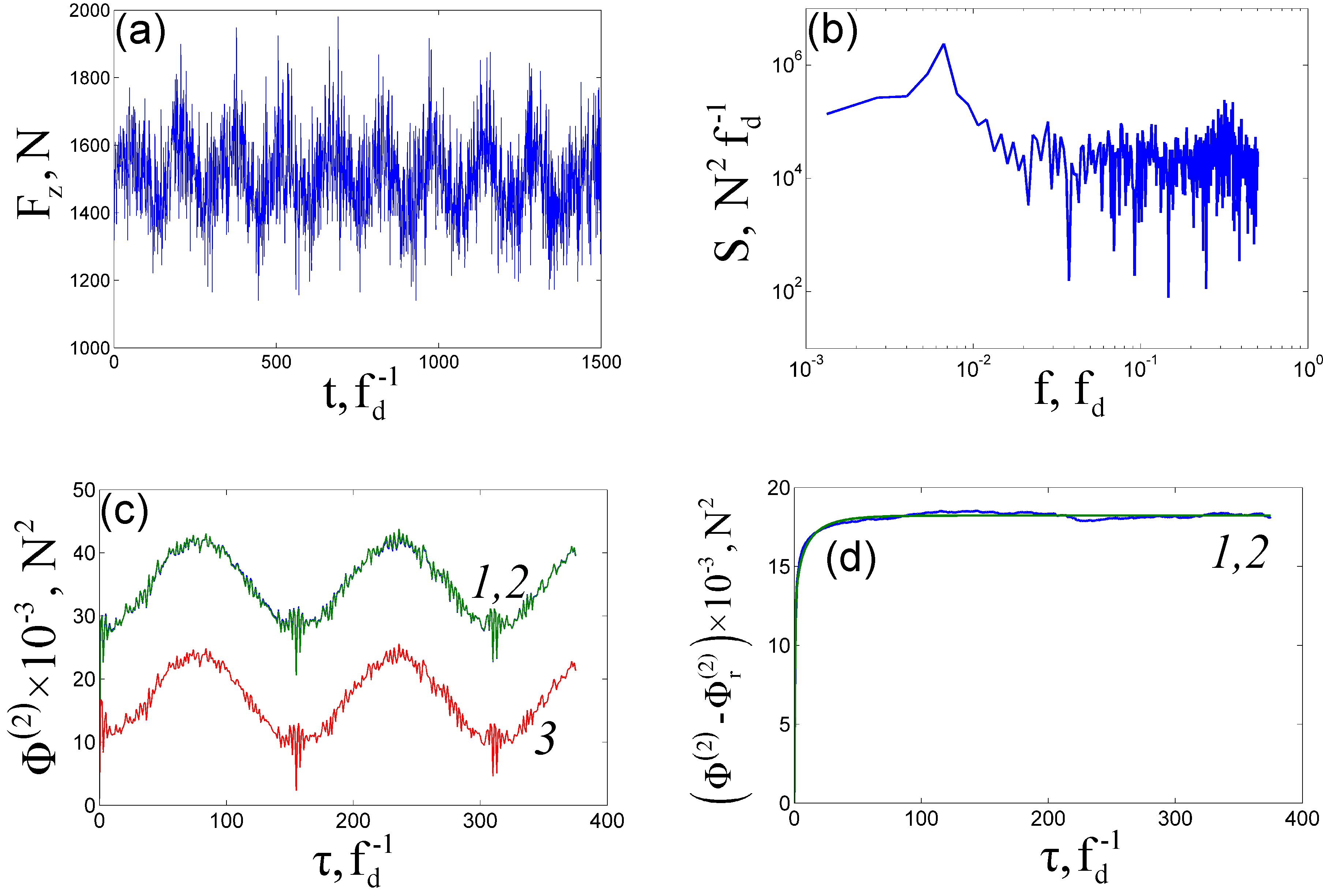}% Here is how to import EPS art
\caption{\label{fig:6} (Color online) FNS parameterization of $F_z$ at $h_d = 2.30$ mm: nomenclature as in Fig. 2.}
\end{center}
\end{figure}

The time series for feed $F_x$, thrust $F_y$, and cutting $F_z$ force components, which are expressed in newtons [N], at $h_d  = 1.00$ mm are plotted in  Figs.~\ref{fig:1}(a),~\ref{fig:2}(a), 
 and~\ref{fig:3}(a), respectively. The corresponding power spectrum estimates on a log-log scale are illustrated in  Figs.~\ref{fig:1}(b),~\ref{fig:2}(b), 
 and~\ref{fig:3}(b). The FNS parameterization was performed for $T$ =1500 $f_d^{-1}$. The structure functions $\Phi^{(2)}(\tau)$ calculated for experimental 
 time series and the sums of resonant components $\Phi_r^{(2)}(\tau)$ and stochastic components $\Phi_s^{(2)}(\tau)$ 
 are shown in Figs.~\ref{fig:1}(c),~\ref{fig:2}(c), and~\ref{fig:3}(c), respectively. The corresponding stochastic components 
 $\Phi_s^{(2)}(\tau)$ are illustrated in Figs.~\ref{fig:1}(d),~\ref{fig:2}(d), and~\ref{fig:3}(d). The values of FNS parameters estimated by the parameterization 
 procedure described in Ref. \cite{Tim12} are listed in Table~\ref{tab:1}. Figures ~\ref{fig:4}-\ref{fig:6} show the same plots as Figs. ~\ref{fig:1}-\ref{fig:3}, but for $h_d$ = 2.30 mm. 

Prior to analyzing the values of FNS parameters for these time series, we checked all nine signals for the quasistationarity criteria described in section 2, which require the existence of a finite value of standard deviation for the stochastic component within the averaging window under study. It can be seen from Figs.~\ref{fig:1}(d),~\ref{fig:2}(d),~\ref{fig:3}(d),~\ref{fig:4}(d),~\ref{fig:5}(d), and~\ref{fig:6}(d) that all of the signals except for $F_y$ at $h_d$ = 2.30 mm meet this requirement. Quantitatively, such quasistationary signals are characterized by the values of $T_1$ smaller than the averaging interval $T$, which is illustrated in Table 1. The nonstationary case of $F_y$ at $h_d$ = 2.30 mm is discussed in detail further in the text.

%fig3

%fig4

\begin{table*}
\renewcommand{\arraystretch}{1.2}
\caption{\label{tab:1}FNS parameterization of force components.}
%\begin{ruledtabular}
%\begin{small}
\scriptsize
\begin{center}
%\begin{tabular}{p{1.5cm}p{1.5cm}p{1.5cm}p{1.5cm}{p{1.5cm}p{1.5cm}p{1.5cm}p{1.5cm}{p{1.5cm}p{1.5cm}p{1.5cm}p{1.5cm}}
\begin{tabular}{ccccccccccc}
& & \\
\hline
 $h_d$ &  $i$ &  $\mu_V$ &  $S_{\max}^{(low)}$ &  $S_{\max}^{(high)}$   &  $\sigma$ &  $H_1$ &  $T_1$ &  $S_s(T_{0}^{-1})$ &  
$n$ & $T_{0}$  \\ ~
 [mm] &  ~  & [N] &  [${\rm N}^2 f_d^{-1}$] &  [${\rm N}^2 f_d^{-1}$] &  [N] &   ~        &  [$f_d^{-1}$]  &  $[{\rm N}^2 
f_d^{-1}$] &  ~ & [$f_d^{-1}$] \\
\hline
1.00 & $x$ & 403 & $5.98 \times 10^5$ & $4.83 \times 10^3$ & 24 & 0.29 & 36 & $2.88 \times 10^3$ & 1.49 & 29 \\
1.75 & $x$ & 595 & $6.92 \times 10^5$ & $3.72 \times 10^3$ & 25 & 0.28 & 38 & $3.37 \times 10^3$ & 1.46 & 31\\
2.30 & $x$ & 743 & $6.11 \times 10^5$ & $6.42 \times 10^4$ & 29 & 0.21 & 46  & $5.55 \times 10^3$ & 1.25 & 38\\
1.00 & $y$ & 581 & $4.94 \times 10^5$ & $5.52 \times 10^4$ & 31 & 0.11 & 27 & $5.89 \times 10^3$ & 0.89 & 34\\
1.75 & $y$ & 741 & $6.52 \times 10^5$ & $8.03 \times 10^4$ & 34 & 0.07 & 108 & $6.78 \times 10^3$ & 0.93 & 36 \\
2.30 & $y$ & 870 & $4.45 \times 10^5$ & $1.91 \times 10^6$ & 66 & 0.06 & $\gg T$   & $1.26 \times 10^4$ & 0.56 & 91\\
1.00 & $z$ & 795 & $1.79 \times 10^6$ & $9.15 \times 10^4$ & 61 & 0.10 & 33 &  $2.47 \times 10^4$ & 0.85 & 40\\
1.75 & $z$ & 1189 & $2.41 \times 10^6$ & $8.51 \times 10^4$ & 74 & 0.09 & 23 &  $3.79 \times 10^4$ & 0.72 & 55\\
2.30 & $z$ & 1502 & $2.39 \times 10^6$ & $2.42 \times 10^5$ & 95 & 0.07 & 27 &  $4.76 \times 10^4$ & 0.64 & 44\\
\hline
\end{tabular}
%\end{small}
%\end{ruledtabular}
\end{center}
\end{table*}

 Figures~\ref{fig:1}(a),~\ref{fig:2}(a), and~\ref{fig:3}(a) and Table~\ref{tab:1} show that for $h_d = 1.00$ mm the mean value is smallest for $F_x$ and largest for $F_z$. Spectral plots
 ~\ref{fig:1}(b),~\ref{fig:2}(b), and~\ref{fig:3}(b) illustrate two competing local maxima: one in the low-frequency (resonant) range and the other in the high-frequency
 (fluctuation) range. The first maximum, in the form of a peak, corresponds to the rotational speed of the spindle (780 rpm or approximately 13 Hz). The second maximum,
 which is obtuse and thus characterizes a frequency band rather than the value at a specific point, is associated with undesired chatter vibrations. The spectral
 values at these maxima are labeled in Table~\ref{tab:1} as $S_{\max}^{(low)}$  and $S_{\max}^{(high)}$ for low-frequency and high-frequency ranges, respectively. For the
 $F_x$ component, the ratio between these spectral values is 124, which implies that the effect of chatter vibrations in this case is insignificant. The graph for difference
 moment (Fig.~\ref{fig:1}(c)) illustrates this fact. For the $F_y$ and $F_z$ components, the ratios are smaller: 9 and 20, respectively, and the plots for difference
 moments (Figs.~\ref{fig:2}(c),~\ref{fig:3}(c)) show discernible fluctuations in the main periodic function corresponding to the spindle rotational speed.

 Figures ~\ref{fig:1}(d),~\ref{fig:2}(d), and~\ref{fig:3}(d) show that the interpolation of the anomalous diffusion type given by Eq.~(5) in Ref. \cite{Tim12} is able to adequately
 describe the stochastic difference moment for all three components. The correlation times (characteristic timescales) $T_1$ and $T_{0}$ are approximately the same while
 the standard deviation $\sigma$ and spectral value $S_s(T_{0}^{-1})$ are smallest for $F_x$ and largest for $F_z$. The values of the flicker-noise parameter $n$ and Hurst
exponent $H_1$  are largest for the $x$-component and smallest for the $z$-component.

Let us see how the values of FNS parameters change when the cutting depth $h_d$ is increased (Table~\ref{tab:1}). 

%fig5

%fig6

%fig7

For the $F_x$ component, the mean, standard deviation $\sigma$, and spectral value $S_s(T_{0}^{-1})$ grow with increasing $h_d$. The timescales $T_1$ and $T_{0}$ are approximately the same for different $h_d$. The main difference is in the value of ratio $S_{\max}^{(low)}$/$S_{\max}^{(high)}$. It is equal to 124, 186, and 10 for $h_d =$ 1.00, 1.75, and 2.30 mm, respectively. This implies that chatter vibrations in the feed force component are insignificant when the cutting depth is 1.00 or 1.75 mm. However, the fluctuations start playing an important role at 2.30 mm, as illustrated by the difference moment plot in Fig.~\ref{fig:4}(c). The values of $n$ and $H_1$, which in this case correspond to the subdiffusion mode, slightly decline as $h_d$ increases. This suggests that the degree of stochastic correlations rises with increasing $h_d$ due to the more pronounced effect of chatter vibrations in the high-frequency range.

For the $F_y$ component, the mean, standard deviation $\sigma$, and spectral value $S_s(T_{0}^{-1})$ also grow with increasing  $h_d$. The value of ratio  $S_{\max}^{(low)}$/$S_{\max}^{(high)}$ is equal to 9, 8, and 0.2 for $h_d =$ 1.00, 1.75, and 2.30 mm, respectively. This implies that chatter vibrations in the thrust force component, which are already significant at $h_d=$ 1.00 and 1.75, play the dominant role at $h_d=2.30$ mm, as illustrated by the difference moment plot in Fig.~\ref{fig:5}(c). This is accompanied by a high value of $T_1$, exceeding $T$ by more than one order of magnitude, which corresponds to a highly nonstationary process producing a large spread in the values of quality parameters for the stainless steel surface formed by turning. Therefore, this mode of stainless steel turning is unacceptable at the cutting depth of 2.30 mm.

For the $F_z$ component, the mean, standard deviation $\sigma$, and spectral value $S_s(T_{0}^{-1})$ also grow with increasing  $h_d$. The timescales $T_1$ and $T_{0}$ are approximately the same. The value of ratio  $S_{\max}^{(low)}$/$S_{\max}^{(high)}$ is equal to 20, 28, and 10 for 1.00, 1.75, and 2.30 mm, respectively. This implies that chatter vibrations in the cutting force component do not change with cutting depth as much as in the cases of $F_x$ and $F_y$. The changes in the values of $n$ and $H_1$ are similar to the case of $F_x$. At the same time, $n<1$ in this case suggests that the stochastic dynamics corresponds to fractional Gaussian noise.

\subsection{Cross-correlations between force components}
\label{3.3}

\begin{table}
\caption{\label{tab:2}Maximum $q_{\max}$ and minimum $q_{\min}$ values of FNS cross-correlation function at different values of cutting depth.}
\begin{center}
%\begin{ruledtabular}
\begin{tabular}{cccc}
& & \\
\hline
$h_d$, mm & components & $q_{\max}$ & $q_{\min}$ \\
\hline
1.00 & $x$-$y$ & 0.75 & -0.73 \\
1.00 & $x$-$z$ & 0.73 & -0.71 \\
1.00 & 	$y$-$z$ & 0.62 & -0.59 \\
1.75 & 	$x$-$y$ & 0.63 & -0.59 \\
1.75 & 	$x$-$z$ & 0.73 & -0.71 \\
1.75 & 	$y$-$z$ & 0.57 & -0.52 \\
2.30 & 	$x$-$y$ & 0.46 & -0.48 \\
2.30 & 	$x$-$z$ & 0.61 & -0.58 \\
2.30 & 	$y$-$z$ & 0.45 & -0.37 \\
\hline
\end{tabular}
%\end{ruledtabular}
\end{center}
\end{table}

\begin{figure}
\includegraphics[width=13.8cm]{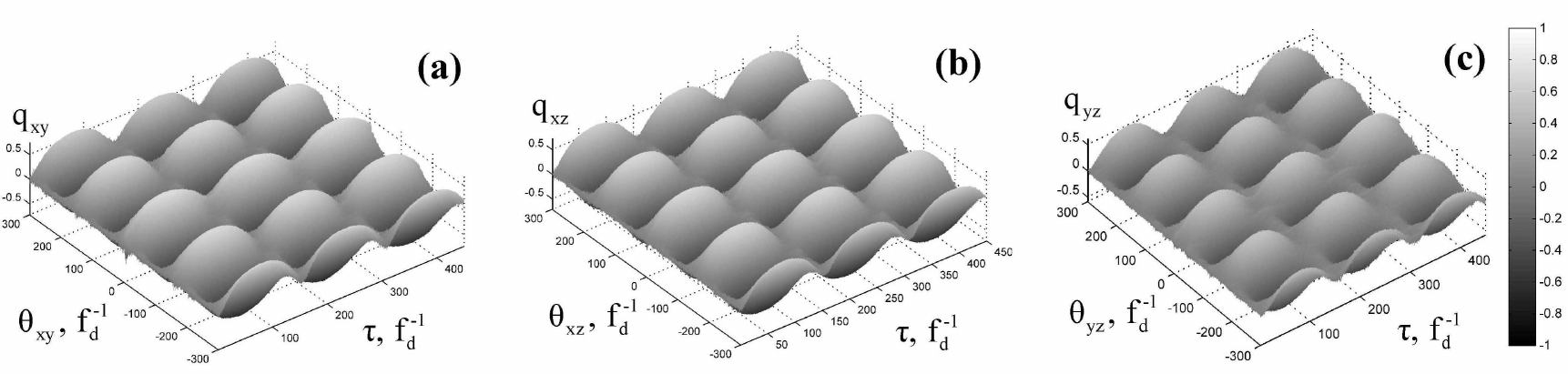}% Here is how to import EPS art
\caption{\label{fig:7} FNS cross-correlation function given by Eq.~(\ref{eq20}) for $x$-$y$ (a), $x$-$z$ (b), and $y$-$z$ (c) force components at $h_d  = 1.00$ mm.}
\end{figure}

\begin{figure}
\includegraphics[width=13.8cm]{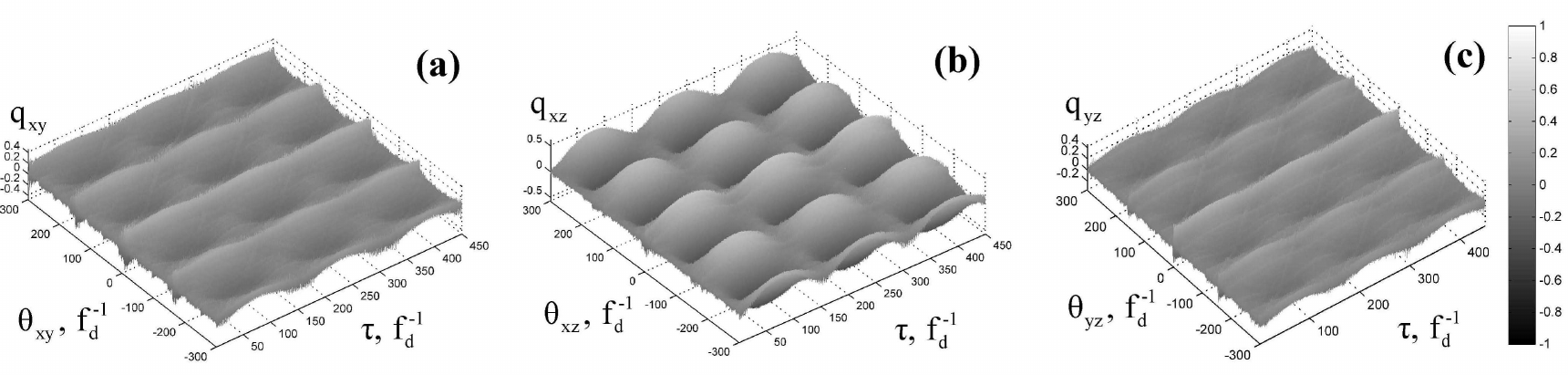}% Here is how to import EPS art
\caption{\label{fig:8} FNS cross-correlation function given by Eq.~(\ref{eq20}) for $x$-$y$ (a), $x$-$z$ (b), and $y$-$z$ (c) force components at $h_d  = 2.30$ mm.}
\end{figure}

Figures~\ref{fig:7} and~\ref{fig:8} illustrate the cross-correlations between different force components at $h_d$ = 1.00 and 2.30 mm, respectively. The cross-correlation functions were calculated at $T =1500 f_d^{-1}$, $\tau_{\max}=450 f_d^{-1}$, and $|\theta_{\max}|=300 f_d^{-1}$. The absolute maximum and minimum values of the cross-correlation function for all three cutting depths are listed in Table~\ref{tab:2}.

%fig8

%fig9

In all six plots (Figs.~\ref{fig:7} and~\ref{fig:8}), we can see the periodicity corresponding to the spindle rotational speed, which suggests that the workpiece is unbalanced. The period is approximately 0.08 sec. The maximum values of correlations (anticorrelations) take place at one of the local maxima (minima) of these "waves". As we saw in the previous section, chatter vibrations correspond to the high-frequency fluctuations appearing on the background of these periodic functions. 

The behavior of cross-correlation functions in Figs.~\ref{fig:7},~\ref{fig:8} can be best explained in terms of frequency and phase synchronization between the force components \cite{Tim10b}. When the chatter vibrations are insignificant (at $h_d$ = 1.00 mm), the frequency and phase for all three components are determined by the rotational speed of the workpiece (forced oscillation), which results in frequency and phase synchronization observed in Fig.~\ref{fig:7}. However, when $h_d$ = 2.30 mm, the chatter vibrations break down the synchronization, and cross-correlations virtually disappear in Figs.~\ref{fig:8}(a),~\ref{fig:8}(c).

It can be seen that the magnitudes of $x$-$y$ and $y$-$z$ cross-correlations decrease with increasing cutting depth (Figs.~\ref{fig:7}(a),~\ref{fig:7}(c),~\ref{fig:8}(a),~\ref{fig:8}(c), Table~\ref{tab:2}). This can be attributed to the fact that the value of  $S_{\max}^{(low)}$/$S_{\max}^{(high)}$  for $F_y$ becomes smaller as cutting depth grows. The $x$-$z$ cross-correlations are virtually the same for $h_d$ = 1.00 and 1.75 mm because the corresponding values of $S_{\max}^{(low)}$/$S_{\max}^{(high)}$ for same components are close in value. On the other hand, the $x$-$z$ cross-correlations for $h_d$ = 2.30 mm are noticeably smaller due to a significant decrease in the value of  $S_{\max}^{(low)}$/$S_{\max}^{(high)}$  for both force components.

\subsection{Comparison with previously reported results}
\label{3.4}

The time series for force components examined in this work were previously studied using statistical, recurrence, multifractal, and wavelet methods \cite{Lit11,Litak2012}. 

The statistical analysis in Ref. \cite{Litak2012} demonstrates that the mean values and fluctuation ranges (standard deviations) for all three components increase with rising $h_d$. This is in agreement with Table 1, although the parameter $\sigma$ in our study refers to the standard deviation only in the stochastic component rather than the whole time series, which includes both regular and noise components. The covariance matrix analysis in the same study implies that the coupling between the force components is generally small, except for the case of correlations between $F_x$ and $F_z$. Our analysis using the FNS cross-correlation function shows that the coupling is generally strong except for the $F_y$ component at the highest cutting depth (see Table 2 and Figs.~\ref{fig:7}-\ref{fig:8}). The discrepancy is caused mainly by the fact that the covariance analysis presented in Ref. \cite{Litak2012} calculates the average values for the whole time series (with no delay between the signals) while the FNS analysis looks at "local" values of the cross-correlation function (considering also the delays between the signals). Since the cross-correlations for the data are usually dominated by the periodic component corresponding to the rotational speed of the spindle, the average value of cross-correlations, representing a sum of both positive and negative values on the "wave", is relatively small. The above implies that this type of covariance analysis does not allow one to capture correctly the effect of synchronization. Recurrence analysis in Ref. \cite{Litak2012} suggests that the vibrations at small cutting depths manifest a collective behavior with the same period of oscillations (spindle rotational speed), which gets progressively disrupted at higher cutting depths. This result is in complete agreement with the cross-correlation analysis presented here (see Table 2 and Figs.~\ref{fig:7}-\ref{fig:8}).

Multifractal analysis for the $F_y$ component performed in Ref. \cite{Lit11} suggests that the values of the H\"{o}lder exponent, which is closely related to the Hurst exponent in this study, are less than 0.5. Same result is reported in this study (Table 1), though we estimated the Hurst exponent after separating out the low-frequency resonant component from the source signal. This can be attributed to the fact that both average H\"{o}lder exponent in Ref. \cite{Lit11} and Hurst component in our study characterize the dynamics of signals mostly at low values of delay $\tau$ or, equivalently, high values of frequency $f$, where the effect of low-frequency resonances is negligible.

Wavelet analysis visually demonstrates that new high-frequency periods emerge in $F_y$ when cutting depth is increased \cite{Lit11}. The result is supported in this work by the (b) plots of Figs.~\ref{fig:1}-\ref{fig:6}. It should be pointed out that in FNS the low-frequency and high-frequency components are separated out, allowing one to quantify the parameters corresponding to high-frequency components (see Table 1), which could be useful for the automated control of product quality.

Various characteristics/techniques of nonlinear time series analysis \cite{Kan04}, including correlation dimension, Lyapunov exponents, Kolmogorov-Sinai entropy, surrogate data tests, singular value decomposition, and many others, have been used to examine the dynamics of steel turning \cite{Gradisek1998a} as well as other manufacturing processes, for instance, laser droplet formation in the welding of electrical contacts \cite{Kre10,Kre11}. The estimation of correlation dimension, Lyapunov exponents, and Kolmogorov-Sinai entropy relying on the Takens' embedding theorem, which is valid only for stationary signals (when applied to real signals, it is assumed that both mean and variance are approximately constant with time at any scale) with a large number of sampling points, is complicated in this case due to the apparent nonstationarity of the signals (as clearly seen in Fig.~\ref{fig:2}a and other figures) and short sampling time interval (only 1,500 points). This implies that attractor reconstruction and reliable estimation of the non-linear characteristics cannot be performed under these conditions. The use of surrogate data tests is hindered by the small number of sampling points. Moreover, the analysis of chaotic (deterministic) features for these data is impeded by the low-frequency resonance (periodic attractor) associated with the regenerative effect, which may lead to false positives when identifying chaotic features, as illustrated in section 3 of Ref. \cite{Pol12}. All of the above suggests that the procedures of nonlinear time series analysis cannot be used to adequately characterize the dynamics of stainless steel turning based on the considered time series. On the other hand, the FNS procedures used in this study have much less restrictive limitations (only wide-sense quasi-stationarity at a specific scale, which is discussed in more detail in Section 2, is needed for the parameterization while no stationarity constraints are applied to the cross-correlation analysis; the resonant component is separated from the overall signal prior to the analysis of the stochastic/random component).

We would like to note that structure functions (\ref{eq4}) and cross-correlation functions (\ref{eq20}) illustrated in Figs.~\ref{fig:1}-\ref{fig:6} [(c) panels] and Figs.~\ref{fig:7}-\ref{fig:8}, respectively, are much more intuitive for analysis than the expressions and plots involved in statistical, recurrence, multifractal, and wavelet methods \cite{Litak2012,Lit11}. The FNS relations can also be used to quantitatively estimate the effect of chatter compared to the main periodicity in force component time series, as demonstrated in Table 2. The above implies that the structure and cross-correlation functions may be applied to control the product quality in manufacturing practice.

\section{CONCLUDING REMARKS}
\label{4}

Our results show that the higher is the cutting depth, the larger are the values of all cutting force components. 
Furthermore, the fluctuations (chatter vibrations) generally get more pronounced when the cutting depth is increased. It should be noted that the relative magnitude of chatter vibrations is different for different force components. The chatter vibrations are the highest at the cutting depth of 2.3 mm for the thrust force component $F_y$ (Fig. 6). The cross-correlation analysis shows that the couplings in $y$-$z$ and $y$-$x$ planes significantly decrease with increasing $h_d$.

The observed effects may be related to both the material properties and dynamics of the rotating workpiece. We suggest that at larger cutting depths $h_d$, the frictional stick-and-slip phenomenon between the cutting tool and the workpiece, which leads to self-sustained vibrations, becomes more important than the regenerative effect (forced oscillation). The frictional stick-and-slip phenomenon is associated with the intermittent sticking and slipping between the contact surfaces of cutting tool and workpiece, resulting in non-linear changes in friction force. It is usually described in terms of chaotic dynamics models \cite{Grabec1988,Warminski2003,Wiercigroch2001}. The regenerative effect representing the behavior where the workpiece geometry from the previous pass influences the dynamics of the following pass may manifest itself in force component time series as an oscillation with a delay corresponding to the period of spindle revolution \cite{Altintas2000,Stepan2001,Fofana2003}. At smaller cutting depths, we observe mostly the regenerative effect, which is illustrated in Figs.~\ref{fig:1}-\ref{fig:3} as the periodicity with a frequency of approximately 13 Hz and in Fig.~\ref{fig:7} as a frequency-phase synchronization. But at larger cutting depths, the high-frequency chatter, possibly driven by a chaotic mechanism \cite{Litak2012}, starts playing a more significant role than the regenerative effect, which is illustrated in Figs.~\ref{fig:4}-\ref{fig:6} and \ref{fig:8}. The suggested frictional mechanism implies that the material properties (stainless steel) play an important role in the examined cutting process.

The above analysis demonstrates that the FNS parameterization procedure, structure function, and cross-correlation function are relatively simple and intuitive analytical tools that may be used to monitor cutting processes and develop new stability maps for nonstandard materials, including stainless steel and composite materials. In particular, the FNS tools may be applied to study the dynamic effects of other important parameters, including the rotational speed, workpiece surface roughness, material non-homogeneity, and cutting tool geometry, on product quality.

\section*{Acknowledgments}
   The partial financial support of Structural Funds in the Operational Programme - 
Innovative Economy (IE OP) financed from the European Regional Development Fund
- Project "Modern material technologies in aerospace
industry", No. POIG.01.01.02-00-015/08-00 is gratefully
acknowledged.

%\end{acknowledgments}

\bibliographystyle{model1-num-names}
\bibliography{PhysicaA}

\end{document}